# Opto-thermal dynamics in whispering-gallery microresonators

Xuefeng Jiang and Lan Yang[*]

*Department of Electrical and System Engineering, Washington University in St. Louis, St. Louis, Missouri 63130, USA*

*\*Corresponding author: yang@seas.wustl.edu*

## Abstract

*Optical whispering-gallery-mode microresonators with ultrahigh quality factors and small mode volumes have played an important role in modern physics. They have been demonstrated as a diverse platform for a wide range of photonics applications, such as nonlinear optics, optomechanics, quantum optics, and information processing. Thermal behaviors induced by power buildup in resonators or environmental perturbations are ubiquitous in high-quality-factor whispering-gallery-mode resonators and have played an important role in their operation for various applications. Here in this review, we discuss the mechanisms of laser field induced thermal nonlinear effects, including thermal bistability and thermal oscillation. With the help of the thermal bistability effect, optothermal spectroscopy and optical non-reciprocity have been demonstrated. On the other hand, by tuning the temperature of the environment, the resonant mode frequency will shift, which could also be used for thermal sensing/tuning applications. Thermal locking technique and thermal imaging mechanisms are discussed briefly. Last, we review some techniques to realize thermal stability in a high-quality-factor resonator system.*

## Introduction

In the last two decades, whispering-gallery-mode (WGM) microresonators[1] have emerged as frontrunners to enable numerous advances in fundamental science and technology developments, including optomechanics, non-Hermitian physics, communications, frequency combs, high performance sensors, and cavity quantum electrodynamics (QED), just to name a few[2–23]. Their capability to trap light in a highly confined volume for a long period of time significantly enhances light-matter interactions and enables high power buildup[24,25]. Consequently, thermal effects and the associated dynamics are ubiquitous in WGM microresonators. For example, thermo-optic nonlinear dynamics[26–29] and thermal instability[30,31] have been observed in various applications. Specifically, temperature fluctuation affects the material refractive index and the size of the resonator, both of which modify the mode distributions and shift the resonant frequencies of WGMs. Stable and continuous operation is critical for many applications, such as bio/chemical sensing[32–42], optomechanics[19,43–45], microlaser[46–51], non-Hermitian physics[15,20], and nonlinear photonics[2,3,60,52–59]. Taking WGM sensing as an example, resonance shifts induced by the target of interests[36] are typically mixed with the thermally induced mode shift. Therefore, various thermal-stability techniques have been developed to suppress the thermal noise for applications of WGM microresonators, such as sensing and metrology, where thermally induced signal fluctuations are undesired. In addition, by making use of the thermal effects, innovative photonic techniques have been developed. For example, the thermal tuning technique has been applied to adjust the resonator frequency in parity-time-symmetric resonator systems[20,61,62] and to accurately measure optothermal properties of resonators[63,64]. Furthermore, thermal sensing applications have also been demonstrated in many WGM microresonators using the thermo-optic effect and/or the thermal expansion effect. Besides, thermal scanning technique has been demonstrated to reduce the noise of frequency comb and optical solitons[65]. Moreover, thermal locking[66,67] and thermal imaging techniques[68–70] have also been developed by taking advantages of thermal instabilities and thermal absorption, respectively, which will find broad applications in sensing, microscopy, and spectroscopy.

Here, in this review, the physical mechanism of the thermal nonlinear effects induced by probe/pump laser fields will first be discussed, including both the thermal bistability in a microresonator with uniform material composition and the thermal oscillation in a hybrid-material microresonator. Furthermore, the optothermal spectroscopy, the thermal relaxation parameter measurement, and optical thermal non-reciprocity will also be introduced as applications of the thermal bistability effect. Secondly, thermo-optic applications, including thermal tuning, thermal scanning, thermal sensing as well as thermal locking and photothermal imaging techniques, will be discussed. Thirdly, we will review some techniques to realize the thermal stability. Last, we offer a brief summary of thermo-optic dynamics and techniques in WGM microresonators.

## Thermal nonlinear effects

Ultrahigh power build-up benefitting from high quality factor ($Q$) and small mode volume ($V$) significantly enhances the absorption induced thermal nonlinearity in WGM microresonators. Thermal bistability behavior typically occurs when scanning across a high $Q$ mode, which is translated into a linewidth broadening/narrowing behavior[27]. In addition, thermal oscillation may arise in a microresonator made of hybrid materials with different temperature coefficients.

### Thermal bistability

In a WGM microresonator, the resonant frequency/wavelength response to changes in temperature is affected by both thermo-optic and thermal expansion effects. The thermo-optic effect (d$n$/d$T$) represents the temperature's dependence of material refractive index, while the thermal expansion effect transfers the temperature fluctuations into the changes of the cavity size. Both the cavity size and the refractive index could affect the resonant frequency according to the WGM resonant condition ($2\pi n(T)R(T)=m\lambda(T)$). The temperature of a WGM resonator could be tuned by an external thermal source or heat generated from the optical absorption. The external thermal source has been widely applied to resonant frequency tuning and thermal sensing, which will be discussed later. Here in this section, we focus on the temperature change induced by the optical field itself. The resonant wavelength shift ($\Delta\lambda$) as a function of temperature change ($\Delta T$) could be expressed as,

$$\Delta\lambda(\Delta T) = \lambda_0 \left( \frac{1}{n}\frac{dn}{dT} + \frac{1}{D}\frac{dD}{dT} \right) \Delta T \tag{1}$$

where $\lambda_0$, $D$ and $n$ are the resonant wavelength, effective diameter and refractive index of the cold cavity, respectively. $dn/dT$ and $(1/D)(dD/dT)$ represent coefficients of thermal refraction and thermal expansion of the cavity.

The temperature dynamics can be described by[27],

$$C_P \Delta\dot{T}(t) = I \frac{1}{\left( \frac{\lambda_p - \lambda_0(1+\alpha\Delta T)}{\delta\lambda/2} \right)^2 + 1} - K\Delta T(t) \tag{2}$$

where $C_p$ and $K$ are the effective heat capacity and the effective thermal conductivity of the cavity material, respectively. $I$ represents the optical power that actually heats the cavity, and $\delta\lambda$ is the mode linewidth of the cold cavity. The first term of Eq. 2 represents the heat generated from the resonant mode, while the second term stands for the heat transfer to the environment.

The transmission spectrum of a silica microresonator appears as a triangular shape and a sharp dip during the wavelength up- and down-scanning processes, known as thermal broadening and narrowing, respectively. Figure 1 shows the transmission, temperature, and resonance shift of a silica toroidal microresonator during the wavelength up- and down-scanning processes.[27] Taking wavelength up-scanning process for example, as the wavelength of the pump laser approaches the resonant wavelength (at $t \sim 2$ ms), the cavity begins to heat up, which red shifts the resonant wavelength, making the up scan a pursuit process between resonant wavelength and scanning pump wavelength, *i.e.*, both the resonant wavelength and the scanning pump wavelength shift in the same direction. Specifically, the pump follows after the moving resonant wavelength and the detuning between them decreases linearly; therefore more pump is coupled into the resonator as the up scanning proceeds. The pursuit process continues until the resonant point is caught up by the pump wavelength ($t \sim 16$ ms). At this point, the thermal absorption is maximal, and the compensation of the heat dissipation to the environment by thermal absorption cannot be maintained anymore. Beyond this point, the resonant state is rapidly lost since the pump laser cannot push the resonant wavelength further. On the other hand, in the case of wavelength down-scanning process, a narrowed resonance lineshape is observed due to an opposite process between the pump and resonant wavelengths, *i.e.*, thermally induced resonance shift and the scanning pump wavelength move in opposite directions, which gives rise to passing through the resonance quickly.

As an application, thermal bistability could be used to scan the temperature of a WGM resonator during the wavelength up-scanning process.[63] There is a linear region in Fig. 1a, where the temperature of the cavity increases linearly over time. This property could be used to thermally scan WGMs with a fixed-wavelength laser (Fig. 2b). Specifically, a high power tunable laser in 1440 nm wavelength band is used to pump and scan the temperature of the microresonator, while another fixed-wavelength laser in 1550 nm wavelength band is used to probe a high $Q$ WGM when the resonant wavelength is thermally scanned. This thermal bistability assisted scanning scheme provides a stable optothermal spectroscopy by removing the environment thermal noise induced spectral perturbations, which could extend WGM applications to any wavelength bands. Furthermore, optothermal spectroscopy has been used to control transmission spectra and optical gains in both erbium-doped[71] and Raman[72] WGM microlasers.

Recently, the thermal relaxation time and effective thermal conductance of a WGM could also be estimated by fitting two nearby optical modes modulated by the thermal effect.[64,73] Specifically, two nearby WGMs of a toroidal microresonator are scanned by a weak probe laser and a strong probe laser, where the weak probe records the wavelength detuning of these two WGMs, and thus the thermal relaxation time could be derived from the change of the detuning in the case of the strong probe. By tuning the strong probe power, the quantified thermal relaxation process could be fitted.

Optical non-reciprocal devices have attracted increasing attention in the past few years.[74] The standard method to realize optical non-reciprocity is through the magneto-optical effect, which requires an external magnetic field. Magnetic-free optical non-reciprocity in microresonators based on thermal nonlinear effects has been proposed by Fan *et al*.[75] Specifically, a silicon microring resonator coupled with two waveguides form an add-drop filter system. The coupling strengths of these two waveguides are distinct due to the different ring-waveguide gaps. When exciting the resonator in counter directions, the resonator undergoes different thermal broadenings in the transmission spectra. If choosing the resonant point as the operating wavelength, the nonreciprocal transmission ratio up to 40 dB could be achieved in this add-drop filter thanks to the thermal nonlinear effect.

**Thermal oscillation**

Hybrid microresonators could result in novel dynamic properties, such as thermal oscillation. Here, we discuss this phenomenon with a polydimethylsiloxane (PDMS) coated microtoroid as an example.[76] Figure 3 shows the experimental results and numerical simulations of dynamic changes in the transmission (Fig. 3b), temperature fluctuations in both silica and PDMS layers (Fig. 3c), and the resonant wavelength in the wavelength up-scanning process (Fig. 3d). There are four thermo-dynamic processes in the transmission spectrum, marked by regions with different colors.

Region I: From point O to A in Fig. 3b, the probe laser is coupled into the microresonator gradually when the probe wavelength ($\lambda_p$) approaches the resonant wavelength ($\lambda_r$). The temperature of the microresonator increases during this process due to the thermal absorption, leading to a refractive index decrease/increase within the mode volumes in PDMS/silica. As shown in Fig. 3c, the temperature increase in the PDMS layer is much larger than that of the silica layer because the absorption coefficient of PDMS is much larger than that of silica in this wavelength band. Thus, the resonance is dominated by the thermal-optic effect of the PDMS layer, showing a blue shift (green curve in Fig. 3d). On the other hand, considering $\lambda_p$ and $\lambda_r$ shift in opposite directions, their detuning ($\Delta\lambda$) decreases rapidly (purple curve in Fig. 3d), which gives rise to a rapid fall in the transmission spectrum. Point A is the exact resonant frequency of the WGM ($\Delta\lambda = 0$), at which point, the transmission decreases to the minimum value.

Region II: From point A to B, temperatures in both layers keep increasing. Resonant wavelength $\lambda_r$ keeps blue shift at the beginning due to the dominance of PDMS. Considering $\lambda_p$ is large than $\lambda_r$ at point A, the transmission increases rapidly with increasing $\Delta\lambda$ (Figs. 3b and 3d). The influence of the thermal-optic effect in PDMS decreases as the optical absorption decreases, and $\lambda_r$ undergoes a transition process from the blue shift to the red shift in this region. Although the resonant wavelength $\lambda_r$ moves in the same direction as the wavelength up-scanning process after the transition point, the speed of the red shift of $\lambda_r$ cannot catch up with that of the wavelength up-scanning, thus the transmission keeps increasing. Note that the speed of the red shift in $\lambda_r$ increases gradually. Finally, at point B, red shift speed of $\lambda_r$ is the same as that of scanning $\lambda_p$. As a result, at point B, both the transmission and detuning $\Delta\lambda$ reaches a local maximum point.

Region III: From point B to C, the thermal-optic effect in silica dominates due to the stable temperature rise in silica section of the mode volume. Therefore, the red shift speed in $\lambda_r$ is much faster than the speed in $\lambda_p$, resulting in a decrease in both $\Delta\lambda$ and transmission. At point C, $\lambda_r$ equals to $\lambda_p$, and thus the transmission value reaches the local minimum a second time because the optical mode is on resonance again.

Region IV: From point C to D, the red shift in $\lambda_r$ increases continually, resulting in an increase of both $\Delta\lambda$ and transmission. Thus the optical absorption induced heat decreases. When the optical absorption is smaller than the dissipation to the environment, the temperature in the mode volume starts to decrease. The temperature decreasing speed in PDMS is faster than the speed in silica because PDMS has a much larger heat dissipation. Thus, the dynamic behavior of $\lambda_r$ is dominated by the thermal-optic effect of PDMS (green curve in Fig. 3d). The transmission increases quickly due to the fast separation of $\lambda_r$ from $\lambda_p$. Point D represents the non-resonant point, where the wavelength detuning is large enough, and thus the transmission returns to the maximum. After passing through point D, a similar cycle repeats the whole process (O-A-B-C-D), which gives rise to the thermal oscillation with a period from O to D.

The thermal oscillations have been experimentally realized in not only microresonators with coating, such as PDMS coated microtoroid[76], PDMS coated microsphere[77], PMMA coated microtoroid[78], etc., but also uncoated microresonators with uniform material composition such as silica microsphere[79,80], ZBLAN (ZrF4-BaF2-LaF3-AlF3-NaF) microsphere[81], millimeter-sized BaF$_2$ disk[82], silicon nitride[83], and lithium-niobate[84,85] microdisk, etc. The mechanisms of thermal dynamics in uncoated resonators are slightly different from the coated ones. They all result from interplay of multiple effects with different influences on the resonant condition. For example, thermal oscillation in a silica microsphere originates from the interplay between Kerr nonlinear effect and the thermo-optic effect near 20 K[80]; thermal oscillation in ZBLAN microsphere is based on the interplay of three effects, including Kerr effect, thermo-optic effect, and thermal expansion[81]; oscillation in BaF$_2$ disk is due to the interactions of positive thermo-optic effect, negative thermoelastic effect, and intrinsic Kerr nonlinearity[82]; and oscillations in Si$_3$N$_4$ and LiNbO$_3$ microdisks are the results of the interplay of two nonlinear effects, *i.e.*, fast thermo-optic nonlinearity and a slow process, such as thermo-mechanical nonlinearity[83], heat dissipation process[84] or photorefractive effect[85]. It should be noted that the frequency of the thermal oscillation is typically on the order of Hz to kHz. However, Luo *et al.* report MHz-level thermal oscillation in a PDMS coated microsphere, which is considered as a result of the competition between the thermo-optic effect and thermal-expansion effect of the PDMS[77].

## Thermo-optic applications

### Thermo-optic tuning

Tuning WGM resonance is critical for many applications, for example, PT symmetry,[20] tunable microlaser,[86–88] optical filter,[89] cavity QED,[90,91] *etc*. Various frequency tuning techniques have been explored, such as thermal,[92,93] pressure/strain techniques,[86–88,91,94–96] electro-optical,[97,98] magnetic-field,[99] electro-thermal,[100] internal aerostatic pressure,[101,102] chemical etching,[103] laser polishing,[104] *etc*. Every technique has its own limitation. Specifically, electro-optical and electro-thermal techniques are only applicable to special materials; pressure/strain and internal aerostatic pressure techniques are suitable for given resonator structures, such as microbubble; etching and laser polishing techniques are disruptive, leading to permanent changes in the physical structures of the resonators. Among all these techniques, thermal tuning is the simplest one and essentially applies to all the resonator structures/materials.

As an example, here we discuss a directly tuning method of a silicon microresonator through the thermo-optic effect introduced by a visible laser diode.[105] As shown in Fig. 4a, the laser diode beam is focused onto the top surface of silicon microresonator. The resonant frequency shift as a function of the laser diode power is shown in Fig. 4b, which demonstrates a tuning rate of 0.0067 cm$^{-1}$/mW. It's worth noting that this direct thermo-optic tuning is a local and noninvasive method. In Fig. 4c, the temperature along the resonator depends on the angle $\theta$ for different pump positions, which affects the overlap of the laser beam with the resonance mode leading to changes in the thermo-optic effect. Note that the local temperature when pumping at $\theta=\pi/2$ is much larger than the other two positions; this is attributed to the elliptic profile of the focused beam. In addition, the directly tuning method promises a much faster response than the traditional thermal tuning methods. The normalized resonator response in Fig. 4d, defined as the normalized product of the pump power and the transmitted signal, is plotted as a function of the modulation frequency with a cut-off frequency of around 10 kHz. In the inset, the phase difference between the laser power and the transmitted light is presented, and the modulation depth of the signal as a function of the modulation frequency is also plotted.

## Thermal scanning for comb and soliton applications

Frequency comb generations as well as optical solitons, originating from cascaded four-wave mixing in WGM microresonator, have attracted increasing interests in the past decade[2,106–114]. The generations of both comb and soliton typically rely on the tunable laser source, which has intrinsic drawbacks. Specifically, the linewidths of the best tunable lasers are usually on the order of 100 kHz, which is much broader than that of the best single-frequency laser. The broad linewidth, as well as the noisy amplitude of the tunable laser used for pumping, limits the performance of the comb. In contrast, a single-frequency laser could provide a much lower noise and narrower linewidth than tunable lasers by reference frequency locking technique. For example, a narrow linewidth of < 40 mHz has been demonstrated in a locked single-frequency laser[115], which could be used as the pump source of comb to reduce the noise on the generated comb lines.

Joshi *et al.* demonstrated the frequency comb and mode-locked soliton with a single-frequency pump laser by thermally scanning the resonant frequency of the microresonator.[65] As shown in Fig. 5a, an integrated platinum resistive microheater is fabricated on top of an oxide-clad $Si_3N_4$ resonator. The resonant frequency could be controlled by tuning the current of the integrated microheater due to the thermo-optic effect induced refractive index change. A single-frequency laser with a linewidth of ~ 1 kHz is used as the pump laser, which is then amplified by an EDFA and coupled into the on-chip bus waveguide through a fibre lens. The transmitted signals are monitored by a fast photoreceiver (>12.5 GHz) and then analyzed by an optical spectrum analyzer, an RF spectrum analyzer, and an oscilloscope. By applying a triangular modulation to the heater current, the scanned transmission spectrum measured by the oscilloscope is shown in Fig. 5b, where the step-like structures marked by arrows indicate the transitions into mode-locked soliton states[116–118]. Furthermore, the recorded optical comb spectrum agrees with the fitted sech2 pulse spectrum (dashed blue curve in Fig. 5c), corresponding to a mode-locked single-soliton state.

## Thermal sensing

WGM microresonator could be utilized as a thermal sensor due to its ultrasensitive response to the surrounding temperature. In a WGM based thermal sensor, the resonant wavelength/frequency is a linear function of both the refraction index and the size of the WGM microresonator, both of which vary with the temperature of the environment due to the thermo-optic and thermal expansion effects, respectively. In principle, cavity material with a larger thermo-optic coefficient and/or thermal expansion coefficient typically leads to a larger frequency shift and thus allows more accurate measurement of temperature.

Thermal sensing experiments have been demonstrated using silica- and silicon-based devices[119–121]. To improve the sensitivity of a thermal sensor, one could utilize microresonator composed of materials with larger thermo-optic coefficients, such as PDMS,[122,123] UV-curable adhesives,[124] lithium niobate,[125] and dye-doped photoresists,[126] which give rise to a much higher sensitivity. For example, PDMS microsphere thermal sensor has demonstrated a sensitivity of 0.245 nm/K.[123] On the other hand, materials with large thermal expansion coefficients, such as silk, have also been used as a WGM thermal sensor. Specifically, a high thermal sensing sensitivity of −1.17 nm/K has been realized in a silk fibroin microtoroid, which is attributed to the large thermal expansion effect.[127]

An alternative experimental design for temperature sensing is the use of microdroplets, which can be made of a variety of materials, such as dye-doped cholesteric liquids, liquid crystals,[128] oils,[129] *etc*. The advantage of microdroplet-based thermal sensor is the ease of integration with conventional microfluidics, as well as diverse choices of materials with relatively high thermal refraction coefficients. Ward *et al.* have also utilized thin-shelled, microbubbles filled with air for temperature sensing.[130,131] Figure 6 summaries the $Q$ factors and sensitivity of some typical WGM thermal sensors working at room temperature. Note that silk microtoroid thermal sensor demonstrates the highest sensitivity (1.17 nm/K), benefiting from the ultrahigh thermal expansion coefficient of the silk material.[127] It's worth mentioning that lithium niobate is an excellent candidate for self-referenced thermal sensing due to its strong thermo-optic birefringence. A preliminary study has demonstrated the self-referenced thermal sensing with an on-chip lithium niobate microdisk resonator with $Q\sim10^5$ by using different thermo-optic responses of ordinary and extraordinary light in the birefringent material.[125] Experimentally, millimeter-sized lithium niobate WGM resonators[132] with $Q$ factor up to $10^8$ as well as chip-based lithium niobate microdisks[133] with $Q$ factor up to $10^7$ have been achieved, which pave the way to develop a high-performance thermal sensor.

## Thermal locking

Pound–Drever–Hall (PDH) locking technique has been widely used to lock optical frequency of a probe/pump laser to a desired resonant frequency for various microresonator applications, such as optomechanics, nonlinear optics, sensing, *etc*.[134] In this method, a feedback voltage signal is applied to the laser head in order to lock the probe laser frequency to the resonant frequency or a frequency with a constant detuning from the resonance. As shown in Fig. 7a, a PDH error signal is generated by measuring the phase difference between the transmitted light and the probe laser.[66] The proportional–integral–derivative (PID) controller could read the laser's offset from cavity resonance, and feed this into a servo loop, which adjusts the frequency by minimizing the PDH error signal, which is linear with the frequency detuning near the resonance, making the servo loop straightforward to manage.

Similarly, the thermally modified transmission spectrum during the wavelength up-scanning process is a nearly linear function of the wavelength detuning, as shown in Figs. 1a and 2b, which could serve as an error signal for locking. Specifically, the linear region of transmission function will be used as an input to the PID controller to stabilize the optical power coupled into the resonator, as shown in Fig. 7b. In this linear region, a tiny pump power increase (decrease) could heat up (cool down) the temperature of the resonator and thus increases (decreases) the frequency detuning by pushing the cavity resonance far away from (close to) the pump laser frequency. As a result, optical absorption will decrease (increase). The whole process forms a negative feedback loop, which enables the thermal locking.[66] It is worth noting that this thermal locking technique is only capable of locking to the resonance slope in the blue detuning region.

However, the external environment perturbations may break the thermal locking status, which makes it hard to achieve long-term locking. Therefore, to solve this problem, McRae *et al.* have proposed another locking mechanism by combining the thermal locking technique with an optical locking technique, which could achieve a long-term locking up to 12 hours.[67] Specifically, the thermal locking technique is utilized to lock the resonant frequency of the microresonator to the probe laser frequency by optimizing the balance between the optical absorption and the thermal dissipation. While an optical feedback locking system induced by some scattering centres is utilized to achieve fast feedback control. It locks the probe laser frequency to resonant frequency by maximizing the constructive interference between the intracavity field and the feedback laser field.

## Photothermal absorption spectroscopy and imaging

Previous studies are based on the thermal dynamics of the whole cavity or materials in the mode volume, while small temperature change in the local environment of microresonators could also be detected by the thermal shift of the resonance, a similar effect used in photothermal absorption spectroscopy.[68] As shown in the left inset of Fig. 8a, high $Q$ WGMs in a microtoroid are excited through a tapered fibre. To perform the photothermal absorption spectroscopy, a second free-space beam is focused onto the surface of the microresonator, where a single gold nanoparticle is deposited. The excitation of the gold nanoparticle gives rise to a local temperature increase due to the photothermal absorption, which red shifts the resonance. The photothermal absorption spectroscopy could be achieved by tuning the pump laser wavelength. PDH locking system could be applied to the probe laser to improve the thermal shift sensitivity down to a single attometer. It should be noted that here an all-glass microtoroid (the right inset of Fig. 8a) is used to perform the photothermal absorption spectroscopy experiment in order to reduce the photothermal background noise of the silicon pillar of a normal silica-on-silicon toroid in the visible region.

On the other hand, the photothermal absorption imaging could also be obtained by scanning the pump beam across the surface of the microresonator, as shown in Figs. 8b and 8c.[68–70] In the experiment, mode shift signals are recorded in real time when scanning the pump laser beam spatially on the top surface of the resonators, which create a thermal shift image. Experimentally, large area thermal mappings of a normal silica toroid on silicon pillar (left) and an all-glass toroid (right) are taken to compare the background noises (Fig. 8b). The background noise of the photothermal image for the all-glass microtoroid is much lower than that of the normal toroid on a silicon pillar at 630 nm wavelength band, which represents a great advantage of the all-glass microtoroid resonator as a platform for photothermal imaging. Furthermore, individual gold nanorods with a size of 40 nm by 80 nm are deposited onto the all-glass microtoroid, which are then imaged by thermal shifts. Experimentally, large-area photothermal imaging is first utilized to find the position of the nanorod on the microtoroid (Fig. 8b). The high-resolution thermal imaging technique is then performed to image individual nanorods, as shown in Fig. 8c, with an SEM image presented in the top inset.

## Thermal stabilization

Thermal stability of the WGM resonance is of great importance in almost all applications, such as laser, optical sensor, optomechanics, nonlinear optics, sensing and imaging, *etc*. Various techniques have been developed to eliminate or reduce thermal noises to realize the thermal stabilization, which will be summarized in this section.

### Thermo-optic coefficient compensation

Microresonators immune to thermal fluctuations have been fabricated by either optimizing the resonator structures/geometries or materials. For example, in a liquid core optical ring resonator, optimization of the resonator's geometry with respect to the surrounding medium can facilitate a reduction of the WGM's thermal sensitivity. As shown in Fig. 9a, the liquid core optical rings with a wall thickness of 1.7 μm for an aluminosilicate microring and 2.6 μm for a fused silica microring demonstrate the minimum susceptibilities to temperature fluctuations.[135]

Hybrid microresonators fabricated by materials with opposite thermo-optic coefficients could also realize thermal stabilization. Theoretically, thermal shift of WGMs in a silica microsphere with a positive thermo-optic coefficient could be compensated by coating a thin layer of soft material with an opposite thermo-optic coefficient onto the surface of the microsphere.[136] He *et al.* demonstrated that a thin layer of PDMS coated on the silica microtoroid resonator could eliminate the thermal shift and broadening due to thermo-optic compensation enabled by the opposite thermo-optic coefficients (Fig. 9b).[137] Similar experiments have also been realized in KD-310 glue[138] and quantum dot[139] coated microspheres (Fig. 9c). In addition, compensation of thermal effect in a microsphere has also been demonstrated by optimizing the glycerol content in surrounding aqueous medium (Fig. 9d).[140] The same stabilization scheme is utilized to characterize the thermal response of various polymer layers as well as protein molecules adsorbed to the resonator's surface.

## Thermal locking stabilization technique

As mentioned previously, optothermal spectroscopy using thermal bistability has been adopted to eliminate the spectral perturbations caused by temperature fluctuations[63]. Here, we introduce another active thermal stabilization technique assisted by thermal locking of an additional laser.[141] Two WGMs with different resonant wavelengths and a similar mode volume were exploited in this approach. As shown in Fig. 10a, two tunable lasers are used; one is working in 1550 nm band as a pump laser, and the other is operating in 1450 nm band, serving as a probe laser. Two WDM couplers are utilized to combine the pump and probe lasers, and then separate the transmitted signals. A fibre taper waveguide is used to couple light into WGMs of a microtoroid. In the experiment, the WGM in 1450 nm wavelength band is first scanned and recorded; the thermal nonlinearity is shown as the red curve in Fig. 10b. Then the pump laser at 1550 nm wavelength band is applied and thermally locked to the linear region of the second high $Q$ mode in 1550 nm wavelength band. Heating induced by 1550 nm laser shifts the probe mode in 1450 nm wavelength band. With the help of this thermal locking stabilization technique, the probe WGM in 1450 nm wavelength band appears as a perfect Lorentzian lineshape, as shown in Fig. 10b (blue curve). In addition, Fig. 10c shows the simulation results, which fits very well with the corresponding experimental data.

## Self-referenced dual-mode stabilization technique

Another scheme to stabilize the resonator temperature is self-referenced dual-mode temperature stabilization technique based on simultaneous read-out of TE and TM WGMs. Specifically, the resonant frequencies of TM and TE WGMs in a microresonator have different temperature coefficients, and thus the self-referenced mode shift difference between them may serve as a perfect temperature sensing signal, which could also be utilized for active stabilization of cavity temperature.[142–145]

Experimentally, a millimeter-sized $MgF_2$ disk resonator is excited by an angle-polished fibre coupler. TE and TM modes are controlled by a fibre-based polarization controller, split by a polarizing beam splitter (PBS), and then monitored by two photodetectors, as shown in Fig. 11a. A data acquisition card is used to collect the signals of the photodetectors, which also provides feedback to an electro-optic modulator (EOM) to achieve the self-referenced dual-mode stabilization process. A pair of TE and TM WGMs with small frequency detuning are shown in Fig. 11b.[142] The average frequency of these two resonances is utilized to locate the centre of the next frequency sweep. While the difference $\Delta f = f_o - f_e$ serves as the feedback error signal of temperature stabilization. Two feedback loops are used to provide temperature stabilization. The heaters in the brass cube are the first feedback loop for long-term stabilization. While an amplitude EOM with high-speed heating effect is used to adjust the optical power, which provides short-term stabilization. Temperature stabilization on the order of a few nano-Kelvin has been achieved (Fig. 11c).

## Conclusion

Herein, we have briefly reviewed the mechanisms, techniques, and applications of thermo-optic dynamics in WGM microresonators. The thermal nonlinear effects, including both thermal bistability and thermal oscillation, have been discussed. Optothermal spectroscopy, thermal relaxation parameter measurement and optical thermal non-reciprocity in microresonators could be achieved with the help of thermal bistability effects. Thermal tuning and thermal sensing are two examples of applications of the thermo-optic effect, which have been explored in various WGM resonator structures. By thermally scanning the resonant frequency of a microresonator, a single-frequency laser with much lower noise and narrower linewidth could be used to reduce the noise of frequency comb and optical soliton. In addition, thermal locking techniques and photothermal imaging implemented in WGM microresonators have also been reviewed. Last, we have introduced some techniques to realize thermal stabilization, which is a critical prerequisite to achieve many high-performance WGM applications.

The research field of thermal-optic dynamics in WGM keeps growing. Looking ahead, more technology development and new applications will benefit from the discoveries in this particular area. For example, PDH locking technique could be exploited for thermal sensing by monitoring the feedback voltage signal[146]. Fast thermal scanning technique may find applications in optomechanics, nonlinear optics, optical trapping, sensing and imaging. Thermal stabilization techniques hold a great potential for ultrastable microlaser by suppressing the temperature fluctuations. Last but not least, resonator assisted photothermal absorption spectroscopy and imaging technique will open up new avenues to study materials and structure properties in nanoscale. It's also worth noting that more opportunities will arise when materials with diverse thermal-optic properties are utilized in the resonators. Similar physics and mechanisms as well as the associated technologies could be explored in other optical resonant structures.

## Acknowledgements

This work was supported by the NSF grant No. EFMA1641109 and No. ECCS1711451, ARO grant No. W911NF1710189.

## Author contributions

All authors contribute to the preparation of the manuscript.

## Conflict of interest

The authors declare that they have no conflict of interest.

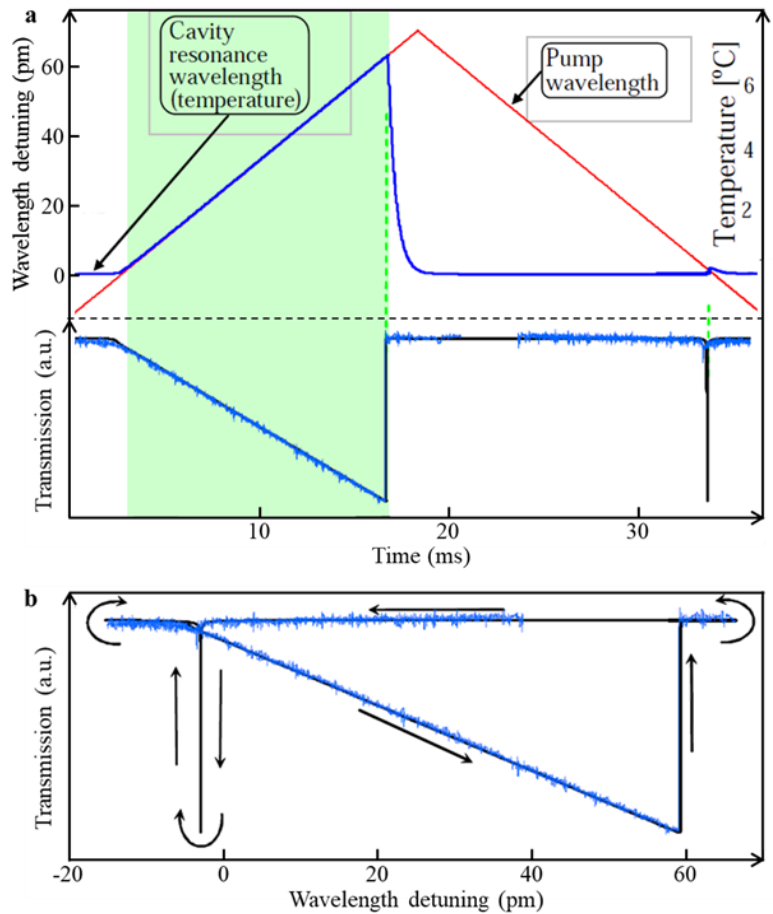

**Fig. 1. Dynamic thermal bistability in a silica microtoroid resonator.**[27] **a** Calculated transmission spectrum, cavity temperature, and resonant wavelength as well as the measured transmission spectrum during the wavelength up- and down-scan processes. **b** Repeating transmission as a function of wavelength detuning in the up-scanning (broadening) and down-scanning (narrowing) processes. Reprinted with permission from ref.[27] [OSA The Optical Society]

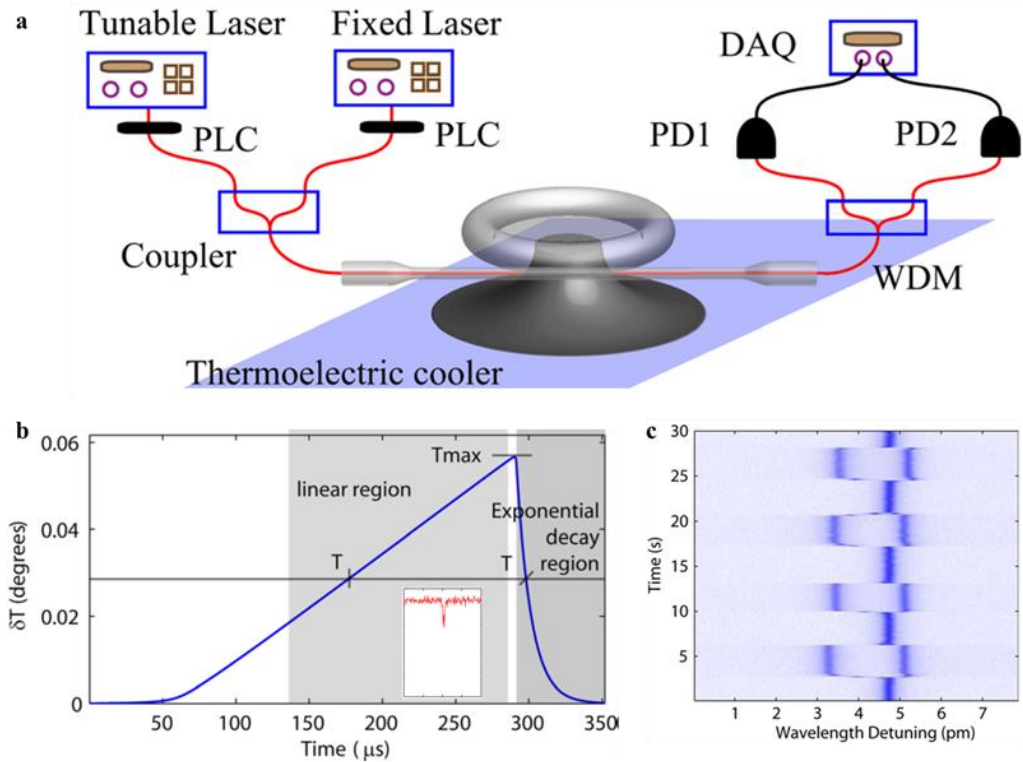

**Fig. 2. a** Schematics of the setup for optothermal spectroscopy. PLC, polarization controller; PD, photodetector; PM, power meter; DAQ, data acquisition; WDM, wavelength division multiplexer. **b** Calculated temperature change of a high *Q* mode during the wavelength up-scanning process. Inset: transmission of a high *Q* mode in 1550 nm wavelength band scanned by a thermal pump WGM in 1440 nm wavelength band.[50] **c** Transmission spectrum frames acquired by optothermal spectroscopy. Here, a nanofibre tip is repeatedly approaching and moving away from the microtoroid.[63] Reprinted with permission from ref.[63] [American Institute of Physics]

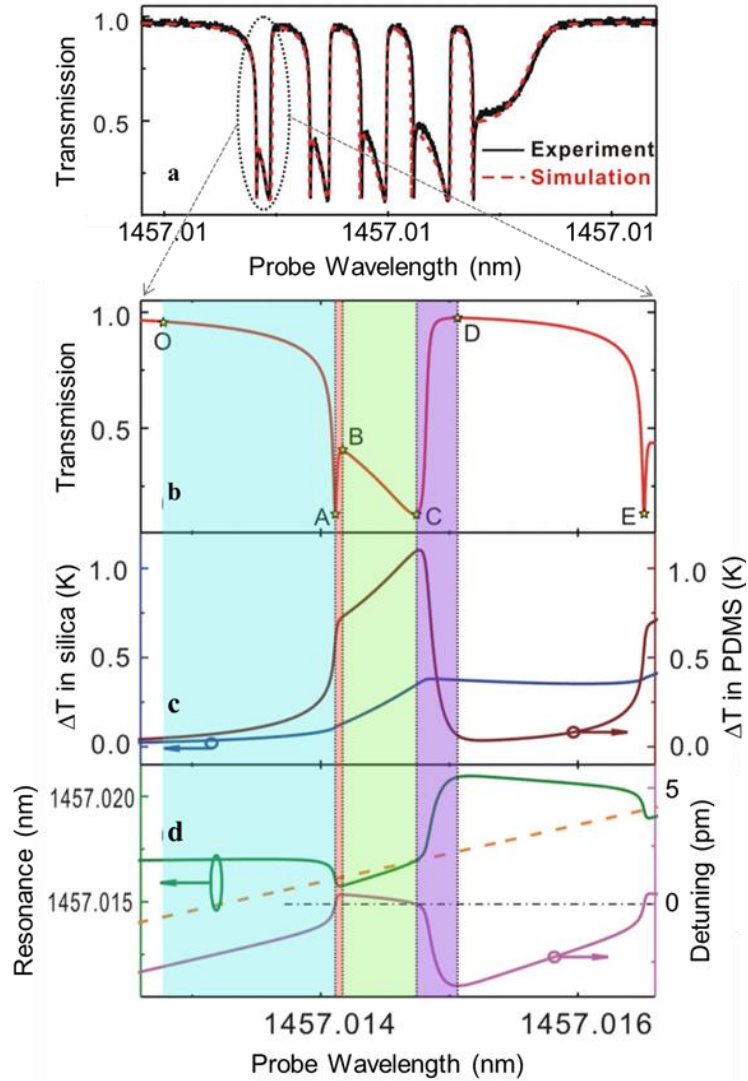

**Fig. 3. Thermal oscillations in a PDMS-coated silica microtoroid during the wavelength up-scanning process. a** Measured and simulated optical transmission spectrum. **b** Simulated transmission oscillation spectrum. **c** Simulated temperature dynamics in both silica and PDMS. **d** Calculated resonant wavelength ($\lambda_r$), probe wavelength ($\lambda_p$), as well as the associated variation in the detuning. Zero detuning is marked by a dot-dashed line (black) in **d**.[76] Reprinted with permission from ref.[76] [OSA The Optical Society]

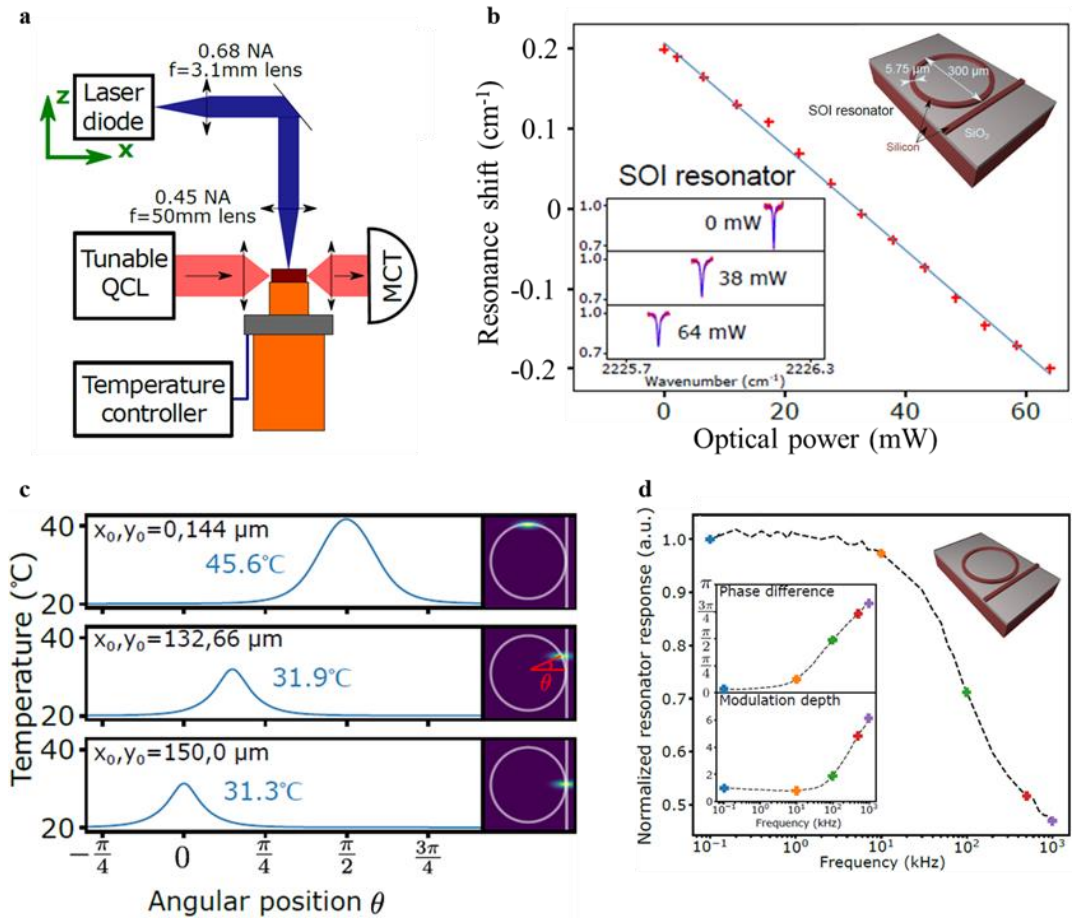

**Fig. 4. Direct thermo-optic tuning of a silicon microresonator. a** Schematic of the setup for direct thermo-optic tuning. **b** The resonant frequency is shifting while adjusting the optical power of the laser diode that serves as the pump to drive the thermo-optic effect. Bottom-left inset: the transmission signal at different optical powers, showing the resonant shift. Top-right inset: schematic of the device. **c** For a focused beam with a dimension of $w_x \sim 40$ μm, $w_y \sim 13$ μm, the temperature profile along the angular position is plotted for three different beam positions. **d** The normalized response of the resonator as a function of the modulation frequency. Inset: the phase difference between the laser power and the detected mid-IR signal and the modulation depth of the mid-IR signal.[105] Reprinted with permission from ref.[105] [OSA The Optical Society]

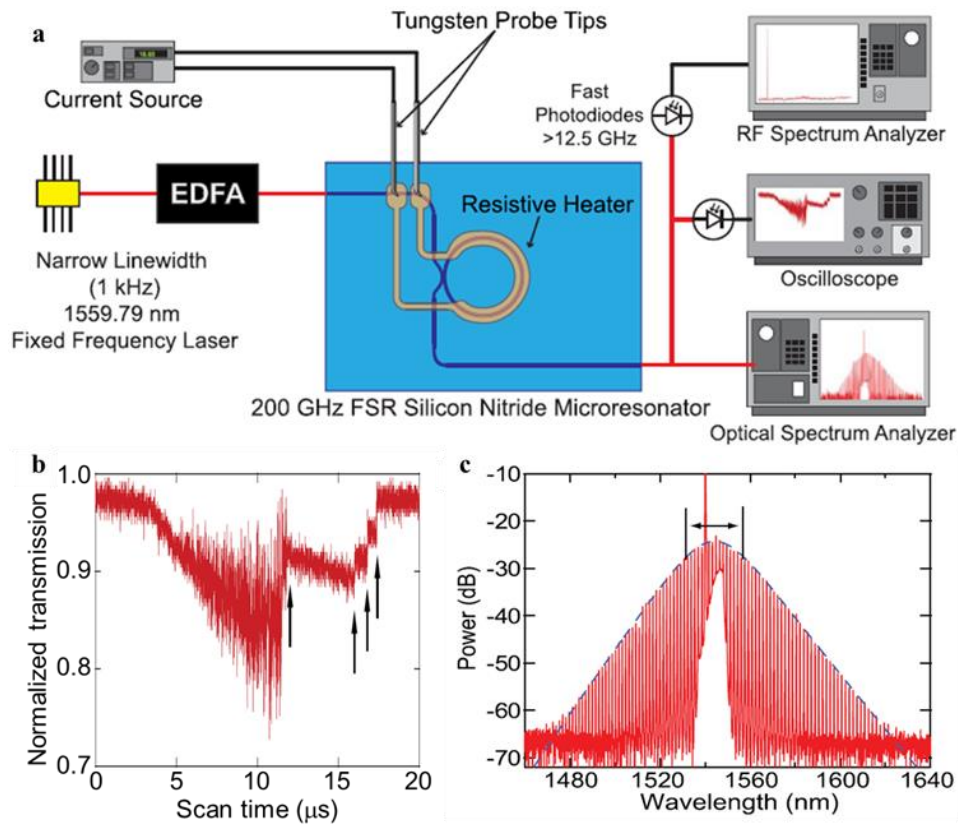

**Fig. 5. a** Schematics of the setup for generation and characterization of optical solitons with thermal scanning method. **b** Oscilloscope trace of the pump transmission as the current on the integrated heater is modulated with a triangular waveform. The steps indicated by the arrows are characteristic of transitions between different multi-soliton states. **c** Measured optical spectrum for a single-soliton mode-locked state with the fitted sech2- pulse spectrum (blue dashed line).[65] Reprinted with permission from ref.[65] [OSA The Optical Society]

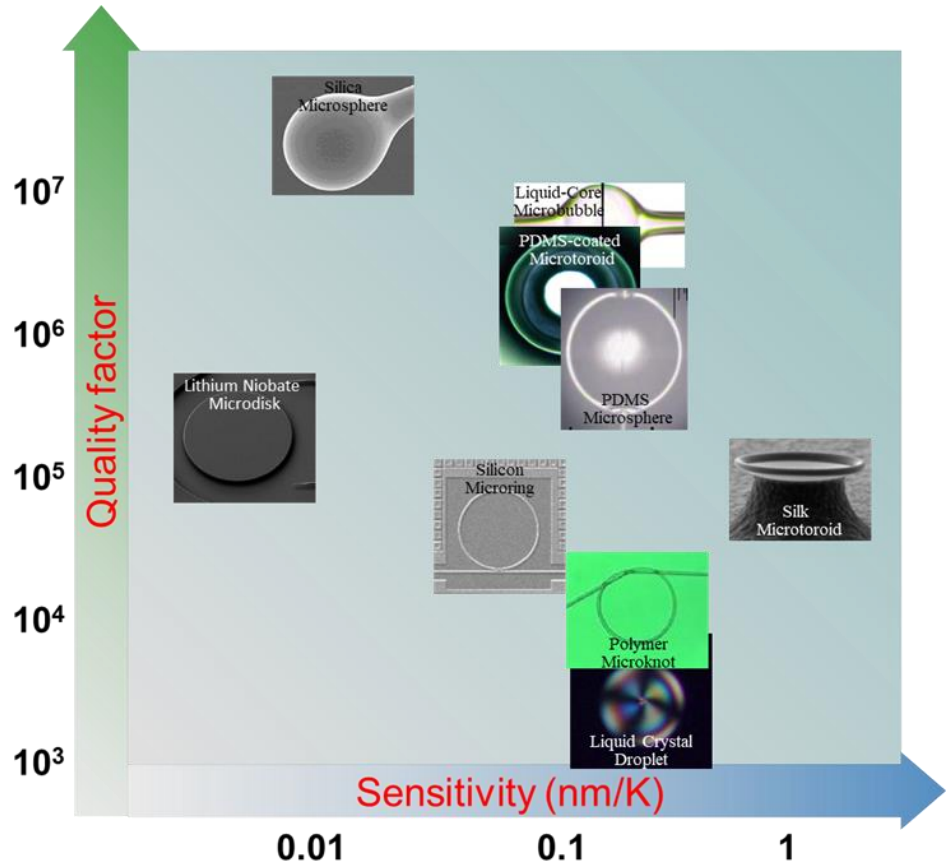

**Fig. 6.** *Q* **factor and sensitivity of WGM thermal sensors made of different materials at room temperature.**

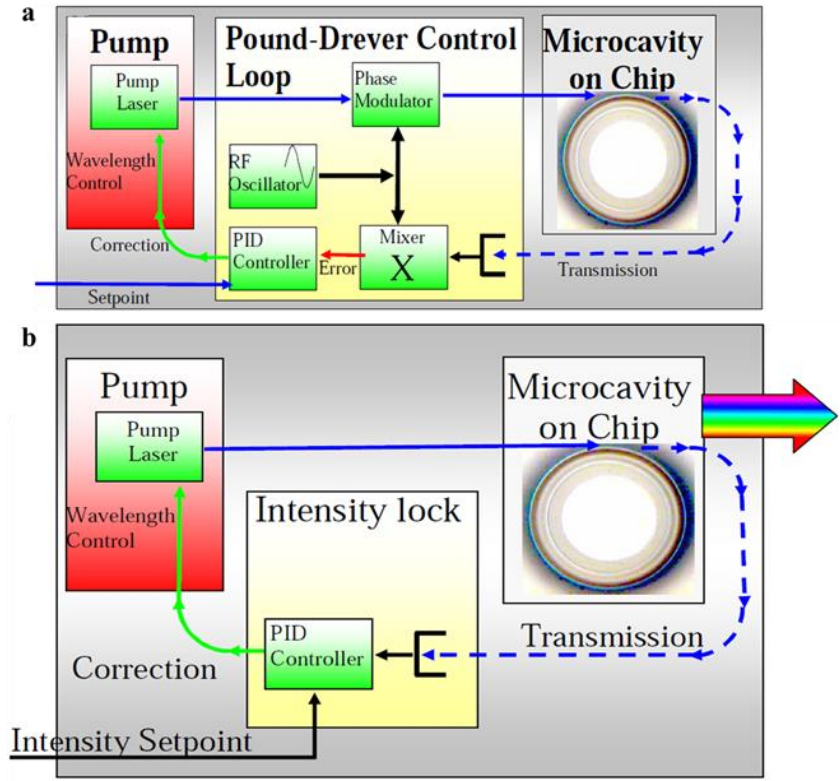

**Fig. 7. Experimental setup of Pound–Drever–Hall (PDH) locking system (a) and thermal locking system (b).**[66] Reprinted with permission from ref.[66] [OSA The Optical Society]

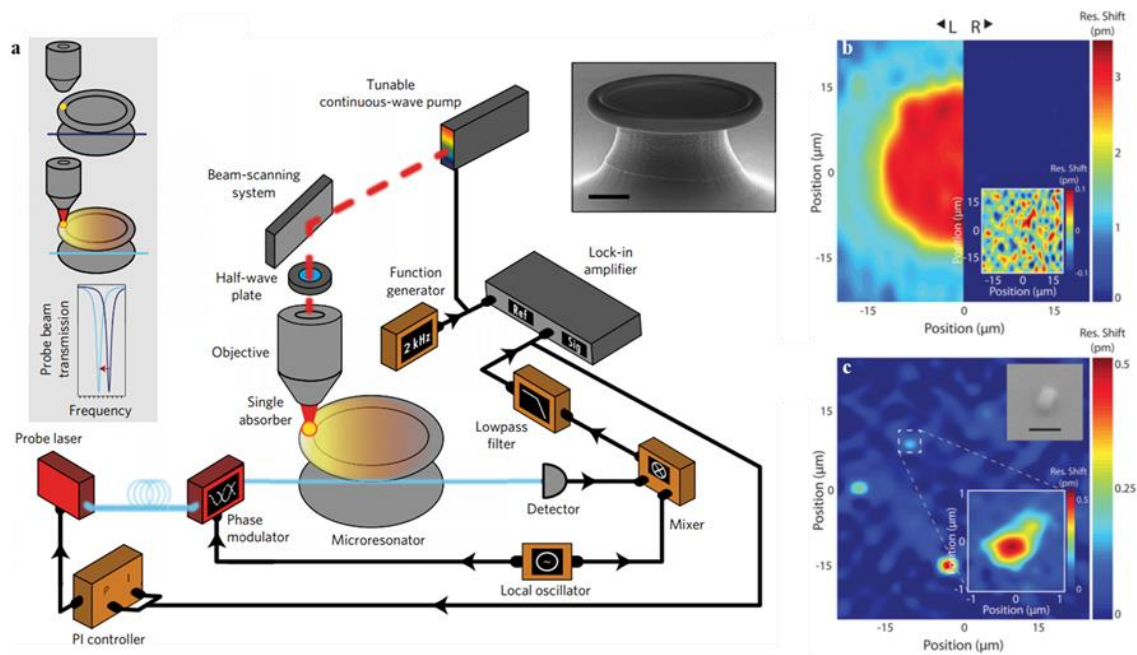

**Fig. 8. a** Schematics of the setup of the photothermal absorption spectroscopy and imaging. Inset (left): diagram of the thermal shift induced by the photothermal absorption of gold nanoparticles. Inset (right): scanning electron micrograph (SEM) image of a silica microtoroid. Scale bar, 10 μm.[68] **b** Large area maps of a silica-on-silicon toroid (left) and an all-glass toroid (right). **c** Photothermal imaging of single gold nanorod on an all-glass microtoroid. Inset (top): SEM of a nanorod with a scale bar of 100 nm. Inset (bottom): high-resolution photothermal imaging of the nanorod.[70] **a** Reprinted with permission from ref.[68] [Nature Publishing Group]; **b,c** Reprinted with permission from ref.[70] [John Wiley & Sons, Inc.].

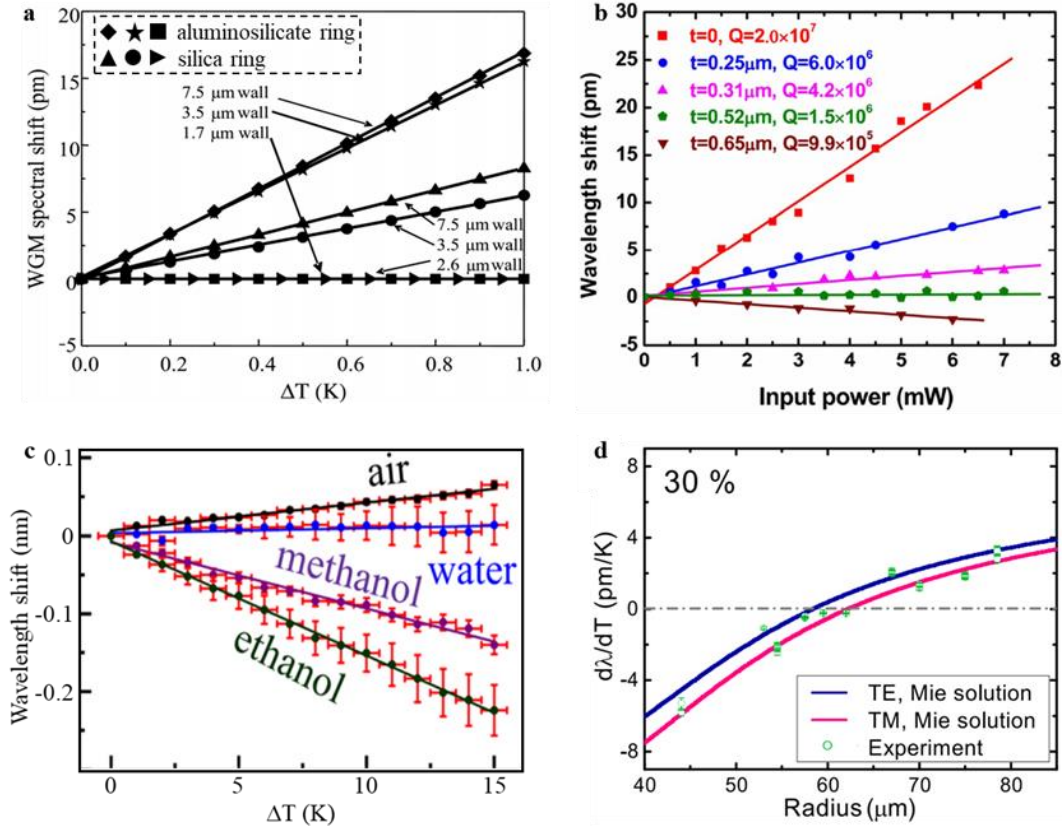

**Fig. 9. Thermo-optic compensation in WGM microresonators. a** Thermal sensitivity reduction by optimizing the geometry of liquid core aluminosilicate and silica optical rings.[135] **b-d** Thermo-optic compensation of WGM microresonators by coating PDMS (**b**)[137], quantum dot (**c**)[139], and glycerol content in surrounding aqueous medium (**d**)[140]. **a** Reprinted with permission from ref.[135] [OSA The Optical Society]. **b** Reprinted with permission from ref.[137] [American Institute of Physics]. **c** Reprinted with permission from ref.[139] [American Institute of Physics]. **d** Reprinted with permission from ref.[140] [American Institute of Physics].

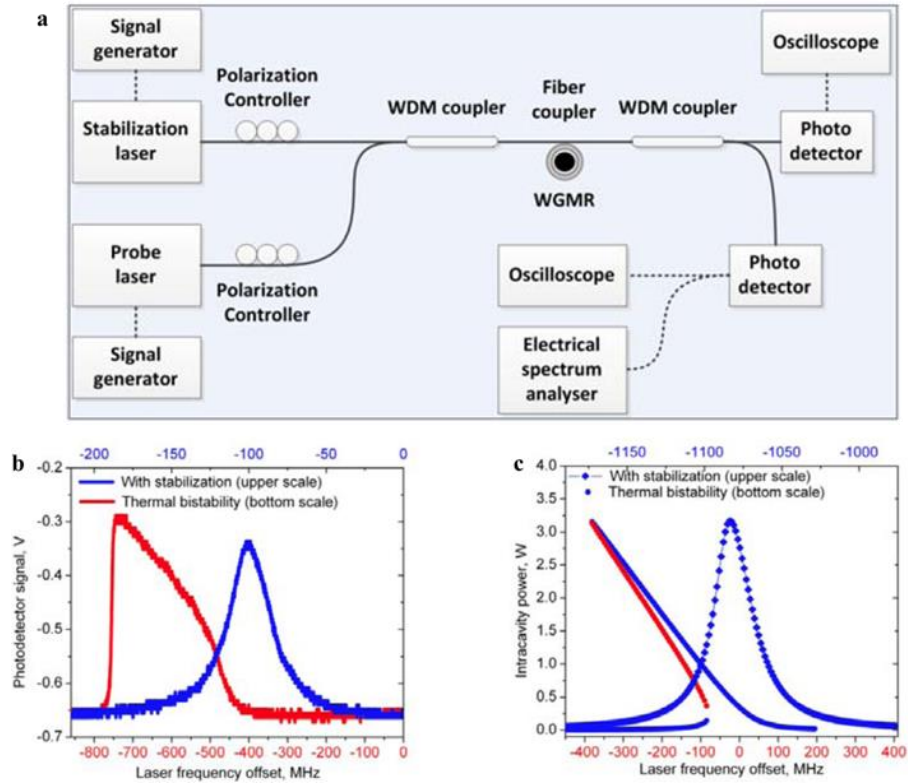

**Fig. 10. a** Schematic of the thermal locking stabilization setup. **b** Experimental spectrum of a high *Q* WGM with (blue) and without (red) thermal locking stabilization. **c** Simulation results of a high *Q* WGM with (blue) and without (red) thermal locking stabilization.[141] Reprinted with permission from ref.[141] [OSA The Optical Society]

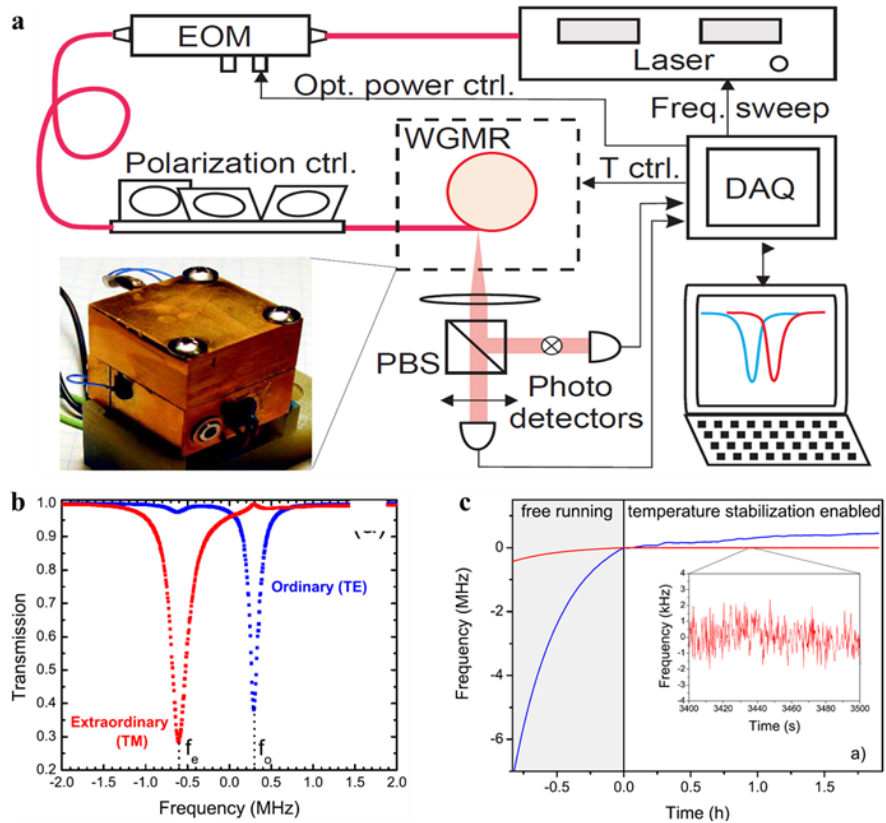

**Fig. 11. a** The experimental setup of a self-referenced dual-mode stabilization technique.[142] **b** A pair of TE and TM WGMs with a small frequency detuning $\Delta f = f_o - f_e$.[142] **c** Long-term absolute resonant frequency and dual-mode frequency are shown as the blue and red curves, respectively, in the case of free-running (t < 0 h) and stabilized (t > 0 h) temperature with dual-mode technique. Inset: Magnification of the frequency fluctuation.[143] Reprinted with permission from ref.[143] [OSA The Optical Society]